\documentclass[conference,a4paper]{IEEEtran}

\usepackage[utf8]{inputenc} 
\usepackage[T1]{fontenc}
\usepackage{url}              % provides \url{...}
\usepackage{cite}             % improves presentation of citations

\usepackage[cmex10]{amsmath}  % Use the [cmex10] option to ensure complicance
\usepackage{mleftright}       % fix to wrong spacing of \left-,
\mleftright                   % \middle- \right-commands 

\usepackage{graphicx}         % provides \includegraphics{...} to
\usepackage{booktabs}         % fixes poor spacing in tables and
 \usepackage[caption=false,font=footnotesize]{subfig}
\usepackage{amsfonts}
\newtheorem{assumption}{Assumption}
\newtheorem{definition}{Definition}

\newtheorem{lemma}{Lemma}

\newtheorem{theorem}{Theorem}

\newcommand{\RomanNum}[1]{\MakeUppercase{\romannumeral #1}}

\newtheorem{innercustomthm}{Theorem}
\newenvironment{customthm}[1]
  {\renewcommand\theinnercustomthm{#1}\innercustomthm}
  {\endinnercustomthm}
\begin{document}

\title{Efficient Alternating Minimization Solvers for Wyner Multi-View Unsupervised Learning} 

%%%%%%

%\author{%
%  \IEEEauthorblockN{Anonymous Authors}
%  \IEEEauthorblockA{%
%    Please do NOT provide authors' names and affiliations\\
%    in the paper submitted for review, but keep this placeholder.\\
%    ISIT23 follows a \textbf{double-blind reviewing policy}.}
%}

%%%%%% Please only add the author names and affiliations for the FINAL
%%%%%% version of the paper, but NOT for the paper submitted for review!
%
%%%%%
%%%%% Single author, or several authors with same affiliation:
 \author{%
   \IEEEauthorblockN{Teng-Hui~Huang and Hesham~El Gamal}
   \IEEEauthorblockA{School of Electrical and Information Engineering, University of Sydney\\
                     NSW, Australia\\
                    \{tenghui.huang, hesham.elgamal\}@sydney.edu.au
                   }
}

\maketitle

%%%%%
%% Abstract: 
%% If your paper is eligible for the student paper award, please add
%% the comment "THIS PAPER IS ELIGIBLE FOR THE STUDENT PAPER
%% AWARD." as a first line in the abstract. 
%% For the final version of the accepted paper, please do not forget
%% to remove this comment!
%%
\begin{abstract}
In this work, we adopt Wyner common information framework for unsupervised multi-view representation learning. Within this framework, we propose two novel formulations that enable the development of computational efficient solvers based on the alternating minimization principle. The first formulation, referred to as the {\em variational form}, enjoys a linearly growing complexity with the number of views and is based on a variational-inference tight surrogate bound coupled with a Lagrangian optimization objective function. The second formulation, i.e., the {\em representational form}, is shown to include known results as special cases. Here, we develop a tailored version from the alternating direction method of multipliers (ADMM) algorithm for solving the resulting non-convex optimization problem. In the two cases, the convergence of the proposed solvers is established in certain relevant regimes. Furthermore, our empirical results demonstrate the effectiveness of the proposed methods as compared with the state-of-the-art solvers. In a nutshell, the proposed solvers offer computational efficiency, theoretical convergence guarantees (local minima), scalable complexity with the number of views, and exceptional accuracy as compared with the state-of-the-art techniques. Our focus here is devoted to the discrete case and our results for continuous distributions are reported elsewhere. 
\end{abstract}

\section{Introduction} 
Multi-view representation learning aims to efficiently extract meaningful information from data collected from different sources for improved performance over individual views. In this growing body of work, it is typically assumed that there exists a shared target across the multiple views. In supervised settings, the common information corresponds to the available labels and can be extracted efficiently through well-established information-theoretic principles~\cite{tishby2000information,9965818,8986754,9154315,alemi2016deep}. On the other hand, the unsupervised settings, where no labels are available for extracting common information, is closely related to Wyner common information framework where an information-theoretic approach is utilized to characterize the common randomness of two (or more) correlated random variables~\cite{wyner1975common}. 

Here, we adopt the Wyner multi-view unsupervised learning paradigm and build upon the recent progress in understanding common information~\cite{kumar2014exact,relaxWynerInfo22,wynerCont16,e21121181,e22020213}. These recent works have made significant contributions in characterizing Wyner common information in several special but practical cases. For example,~\cite{kumar2014exact} introduces the notion of ``exact common information'' which can be viewed as a deterministic variant of the Wyner common information. Moreover, in~\cite{5766249,wynerCont16}, the connection to lossy source coding and an explicit expression for the case of two correlated Gaussian sources are found. Most relevant to our work, a relaxed variant of Wyner common information formulation is introduced in~\cite{relaxWynerInfo22,9834546} and a characterization of the generalized vector Gaussian sources case. 

To the best of our knowledge, only few works have considered the development of computationally efficient algorithms for solving Wyner common information optimization problem. This comes in contrast to the supervised counterpart where several efficient off-the-shelf solvers are available~\cite{alemi2016deep,9518141,9154315,8986754}. More specifically, in the discrete settings, a heuristic gradient-descent-based algorithm was proposed in~\cite{disGrad21} for computing the relaxed Wyner common information, focusing on the comparison to canonical correlation analysis but no convergence guarantee was reported. The focus of our work here is the development of computationally efficient learning algorithms (solvers) for the Wyner multi-view learning in unsupervised setting. Towards this end, we identify two new extensions. The first is referred to as the \textit{Variational} form and is shown to enjoy a linear complexity growth rate with the the number of views. This provides a significant advantage over commonly used methods whose complexity grows exponentially. For this form, we use tools from variational inference to devise an alternating minimization solver targeting a tight surrogate upper bound of our objective function. The second form is referred to as the \textit{Representation} form which includes the two-source relaxed Wyner common information as a special case. We further devise a new alternating direction method of multipliers (ADMM)-based solver for the resulting non-convex optimization problem. We provide convergence guarantees (local minima) for the proposed solvers in certain relevant regimes and, empirically, we show that our solvers can outperform the state-of-the-art in terms of clustering accuracy of a synthetic dataset. Finally, we would like to remark that variational inference was applied to the problem of unsupervised learning in~\cite{e22020213} but the focus of this work was limited to single-view learning context with no straightforward generalization to the multi-view case. 

\paragraph*{Notation}
The upper case letter $Z$ denotes random variables and lower case $z\in\mathcal{Z}$ for realizations (the calligraphic $\mathcal{Z}$ for alphabets). The cardinality of a random variable is denoted as $|\mathcal{Z}|$. The letter $V$ refers to the number of views, and the index set is defined as $[V]:=\{1,\cdots,V\}$. For observations of all views, we define $X^V:=(X_1,\cdots,X_V)$, where for each element, the subscript denotes the view index. $P(X^V)$ denotes the joint distribution of all view observations whereas the subscript $\theta$ in $P_\theta(X^V)$ indicates the parameterized distribution. $\Omega$ denotes a feasible set which might come with extra notation in the subscript for clarity.

\section{The Variational Form}
In multi-view unsupervised learning, Wyner common information can be used as the basis for extracting the {\em shared information} between the multiple views. More precisely, representation $Z^*$ is defined by:

\begin{IEEEeqnarray}{rCl}
    Z^*=&\underset{P_\theta(Z,X^V)}{\arg\min} &\,I_\theta(X^V;Z),\nonumber\\
    &\text{subject to}&\,P(X^V)=\sum_{z\in\mathcal{Z}}P_\theta(z)\prod_{i=1}^VP_\theta(X_i|z).\IEEEeqnarraynumspace\IEEEyesnumber\label{eq:prob_gen_form}
\end{IEEEeqnarray}
It is important to note that in unsupervised learning no labels are available and one can only have access to $P(X^V)$ through the available observations. The problem \eqref{eq:prob_gen_form} is difficult to solve mainly due to the non-convex nature of the feasible solution set of $P(Z|X_1,...,X_V)$. Here we address this complexity problem by borrowing tools from the theory of variational inference~\cite{8588399,blei2017variational}. First, we define a parameterized set of distributions $P_\theta (Z)$, $\{P_\theta (X_i|Z)\}_{i=1}^V$, and recast parameter set of \eqref{eq:prob_gen_form} from $P_\theta(Z,X^V)$ to $\Theta:=\{P_\theta(Z),\{P_\theta(X_i|Z)\}_{i=1}^V\}$.

The second step is to extend the feasible set by allowing $P(X^V)$ and $P_\theta(X^V):=\sum_{z\in\mathcal{Z}}P_\theta(z)\prod_{i=1}^VP(X_i|z)$ to be distinct distributions which should fall within a bounded divergence.
%\begin{equation}\label{eq:gen_dkl_feat_relax}
%    D_{KL}\left[P(X^V)\parallel P_\theta(X^V)\right]\leq \xi,
%\end{equation}
 These two steps now extend \eqref{eq:prob_gen_form} to the following optimization problem: 
\begin{IEEEeqnarray}{rCl}
    &\underset{\theta\in\Theta }{\text{minimize}} &\,I_\theta (X^V;Z),\nonumber\\
    &\text{subject to}&\, D_{KL}\left[P(X^V)\parallel P_\theta(X^V)\right]\leq \xi ,\IEEEeqnarraynumspace\IEEEyesnumber\label{eq:prob_variational_form_2}
\end{IEEEeqnarray}
where $D_{KL}[\mu\parallel\nu]$ denotes the Kullback-Leibler (KL) divergence between two measures $\mu,\nu$ and $\xi>0$ is a threshold. In the third step, we develop the following surrogate upper bound on $I_\theta(X^V;Z)$:
\begin{lemma}\label{lemma:gen_vi_bound}
    Let $P_\theta(X^V):=\sum_{z\in\mathcal{Z}}P_\theta(z)\prod_{i=1}^VP_\theta(X_i|z)$ with $V\in\mathbb{N}\backslash\{1\}$ and $P(X^V)$ is known, then:
    \begin{IEEEeqnarray*}{rCl}
        I_\theta(X^V;Z)\leq -\sum_{i=1}^VH_\theta(X_i|Z)-\mathbb{E}_{X^V;\theta}\left[\log P(X^V)\right].
    \end{IEEEeqnarray*}
\end{lemma}
\begin{IEEEproof}
    See Appendix \ref{appendix:pf_gen_vi_bound}.
    %See Appendix A of~\cite{huang2023efficient}. 
\end{IEEEproof}

The bound is tight when $P_\theta(X^V)=P(X^V)$. In the final step, we combine the upper bound in Lemma \ref{lemma:gen_vi_bound} over $P_\theta(X^V)$ with the extended feasible set \eqref{eq:prob_variational_form_2} through a Lagrange multiplier, giving the overall Lagrangian:
\begin{IEEEeqnarray}{rCl}
    \mathcal{L}_{V;\theta}:&=&-\sum_{i=1}^VH_\theta(X_i|Z)-\mathbb{E}_{X^V;\theta}\left[\log{P(X^V)}\right]\nonumber\\
    &&+\beta D_{KL}[P(X^V)\parallel P_\theta(X^V)],\IEEEeqnarraynumspace\IEEEyesnumber\label{eq:nv_vi_lag}
\end{IEEEeqnarray}
where the scalar $\beta>0$ is a multiplier. The key advantage of this~\textit{variational} form is that it always provides solutions that satisfy the factorization condition $P_\theta(X^V|Z)=\prod_{i=1}^VP_\theta(X_i|Z)$. This results in linear complexity growth $\mathcal{O}(V)$, defined as the number of parameters to optimize with w.r.t. the number of views. This is due to that the parameter set is $\Theta$. Compared to existing methods in literature~\cite{disGrad21} the complexity is reduced from $\mathcal{O}(|\mathcal{Z}||\mathcal{X}|^V)$ to $\mathcal{O}(V|\mathcal{Z}||\mathcal{X}|)$, letting $|\mathcal{X}_i|=|\mathcal{X}|,\forall i\in[V]$.

For simplicity of illustration, we further simplify the relaxed problem by considering uniformly distributed $P_\theta(Z)=1/|\mathcal{Z}|$, reducing the parameter set to $\{P_\theta(X_i|Z)\}_{i=1}^V$ (this assumption can be easily relaxed without affecting the results). Additionally, we treat each fixed $\beta$ as a different optimization problem and sweep through a range of $\beta$ with random initialization.

Our proposed alternating minimization solver, denoted as \textit{Solver} \RomanNum{1}, now follows:
 \begin{IEEEeqnarray}{rCl}
     P_{x_i|z}^{k+1}:=
     \underset{P_{x_i|z}\in\Omega_{i}}{\arg\min}\, \mathcal{L}_{V;\theta}(\{P^{k+1}_{x_j|z}\}_{j<i},P_{x_i|z},\{P^k_{x_l|z}\}_{l>i}),\IEEEeqnarraynumspace\IEEEyesnumber\label{eq:vi_alg}
 \end{IEEEeqnarray}
 where $k$ denotes the iteration counter; $\Omega_i$ is the compound simplex for $i^{th}$ view, formed by cascading the probability simplex for each $z\in\mathcal{Z}$; We define $\{\cdot\}_{<1}=\{\cdot\}_{>V}=\emptyset$. \textit{Solver} \RomanNum{1} updates the vectorized conditional probability $P_{x_i|z}:=P_\theta(X_i|Z)$ (stack along each realization of $z\in\mathcal{Z}$) through \eqref{eq:vi_alg} iteratively until convergence. \textbf{The convergence is assured to a local minima} since by updating a single $P_{x_i|z}$ for each step \eqref{eq:nv_vi_lag} is a convex function of $P_{x_i|z}$, but is non-convex w.r.t. $\{P_{x_i|z}\}_{i=1}^V$~\cite{nesterov2003introductory,NesterovYurii2018Loco,BertsekasDimitriP.1999Np,boyd2004convex}. We refer to Appendix \ref{appendix:conv_solver1} for details.

\section{The Representation Form}
In this approach, we extend the factorization condition using the conditional independence, while fixing the representation condition $P(X^V)=\sum_{z\in\mathcal{Z}}P(X^V)P_\theta(z|X^V)$, which results in the following optimization problem:
\begin{IEEEeqnarray}{rCl}
    &\underset{P(Z|X^V)}{\text{minimize}}&\,I(X^V;Z),\nonumber\\
    &\text{subject to}&\, I(X_S;X_{S^c}|Z)\leq R_{S},\quad\forall S\subset [V]\IEEEeqnarraynumspace\IEEEyesnumber\label{eq:rep_form_gen_problem}
\end{IEEEeqnarray}
where $I(X;Y|Z)$ denotes the mutual information of two random variables $X,Y$ conditioned on $Z$~\cite{cover1999elements} and we drop the subscript $\theta$ in this part with a slight abuse of notation; $S$ is a partition of $[V]$, i.e., $S\subset{[V]},S\cap S^c=\emptyset,S\cup S^c=[V]$ and $R_S\geq 0$ are the control thresholds.
For illustration, let's start with two views; namely $X_1$ and $X_2$ with a known joint distribution $P(X_1,X_2)$. The problem \eqref{eq:rep_form_gen_problem} reduces to:
\begin{align}\label{eq1}
    Z^*=\underset{I(X_1;X_2|Z)\leq \eta}{\arg\min} I(X_1,X_2;Z),
\end{align}
where $\eta>0$ is a control threshold. 
%Then the above leads to the Lagrangian formulation
%\begin{align}\label{eq1-lag}
%    Z^*=\underset{P(Z|X_1,X_2)}{\arg\min} I(X_1,X_2;Z)+\gamma I(X_1;X_2|Z).
%\end{align}

The intuition behind \eqref{eq1} is that each view is composed of common information shared between the two views which is corrupted by two independent noise operators for the two views. By seeking the minimum mutual information representation that satisfies the condition $I(X_1,X_2|Z)\leq \eta$, the common information can be extracted as a function of the threshold $\eta$ which quantifies the allowed correlation between the two views conditioned on the common representation (i.e., residual correlation). If we set $\eta=0$ then the common representation reduces to the auxiliary random variable in the Wyner formulation~\cite{relaxWynerInfo22,wyner1975common}. Note that the special case \eqref{eq1} is also known as the relaxed common information~\cite{relaxWynerInfo22,disGrad21}.

By definition, $I(X_1;X_2|Z)$ can be expressed as:
\begin{IEEEeqnarray}{rCl}
    I(X_1;X_2|Z)&=&I(X_1,X_2;Z)\nonumber\\
    &&-\>I(X_1;Z)-I(X_2;Z)+I(X_1;X_2).\IEEEeqnarraynumspace\IEEEyesnumber\label{eq:condmi_expand}
\end{IEEEeqnarray}
Using the above \eqref{eq:condmi_expand}, along with the known distribution $P(X_1,X_2)$, the problem stated in \eqref{eq1} can be rewritten as:
\begin{IEEEeqnarray}{rCl}
    Z^*=\underset{\sum_{i=1}^2I(X_i;Z)\geq \eta'}{\arg\min}\, I(X_1,X_2;Z),\IEEEeqnarraynumspace\IEEEyesnumber\label{eq:repform_compete}
\end{IEEEeqnarray}
where $\eta'>0$ is a control threshold and $P_\theta(Z|X_1,X_2)$ is the variable. This shows that there are two competing objectives, minimizing $I(X_1,X_2;Z)$ and maximizing $I(X_1;Z)+I(X_2;Z)$. In the following, we develop an ADMM-based solver for \eqref{eq:repform_compete}~\cite{boyd2011distributed}. Furthermore, we will establish the convergence of the proposed solver, building on recent results for non-convex optimization~\cite{attouch2009convergence,attouch2010proximal,li_pong_2016,bolte2018first,bolte2018nonconvex,9518141,huang2022admm}. 

In our context, the original Lagrangian used in ADMM solvers is transformed into the following augmented Lagrangian \cite{boyd2004convex,boyd2011distributed,nesterov2003introductory,BertsekasDimitriP.1999Np}:
\begin{equation}\label{eq:aug_lag}
    \mathcal{L}_c(p,\nu,q):=F(p)+G(q)+\langle \nu,p-q \rangle+\frac{c}{2}\lVert p-q\rVert^2,
\end{equation}
where $\nu$ is the dual variables; $c>0$ is a penalty parameter. Then $\mathcal{L}_c$ is minimized through the following steps iteratively, denoted as \textit{Solver} \RomanNum{2}:
\begin{subequations}\label{eq:alg_admm_sol}
\begin{align}
    p^{k+1}:=&\underset{p\in\Omega_p}{\arg\min}\, \mathcal{L}_c(p,\nu^k,q^k),\\
    \nu^{k+1}:=&\nu^k+c\left[p^{k+1}-q^k\right],\\
    q^{k+1}:=&\underset{q\in\Omega_q}{\arg\min}\, \mathcal{L}_c(p^{k+1},\nu^{k+1},q),
\end{align}
\end{subequations}
where $k$ is the iteration counter. Our \textit{Representation} form can now be mapped into the augmented Lagrangian \eqref{eq:aug_lag} as follows. First, we observe that the Lagrangian of the problem \eqref{eq:repform_compete} can be rewritten as a combination of Shannon entropy and conditional entropy functions \cite{cover1999elements}:
\begin{IEEEeqnarray}{rCl}
    \mathcal{L}_{\gamma}:&=&(1-\gamma)H(Z)-(1+\gamma)H(Z|X_1,X_2)\nonumber\\
    &&+\gamma H(Z|X_1)+\gamma H(Z|X_2),\IEEEeqnarraynumspace\IEEEyesnumber\label{eq:mv_sto_lag}
\end{IEEEeqnarray}
where $\gamma>0$ is the Lagrange multiplier. We then define the following primal and augmented variables pair:
\begin{equation}\label{eq:var_pq}
        p:=-\log{p_{z|x_1,x_2}},\quad q:=-\log{p_{z'|x_1,x_2}},
\end{equation}
where $p_{z|x_1,x_2}$ is the vector form of the conditional probability $P_\theta(Z|X_1,X_2)$, obtained by stacking the conditional probability mass along each view-specific observation $x_i\in\mathcal{X}_i,i\in[V]$. This assignment facilitates the implementation of the solver (e.g., establishing smoothness conditions and simplex projection), the details are included in Appendix \ref{appendix:conv_prelim}.

Next, the two sub-objectives are assigned as follows:
\begin{IEEEeqnarray}{rCl}
        F_\gamma(p)&:=&-(1+\gamma)H(Z|X_1,X_2),\nonumber\\
        G_\gamma(q)&:=&(1-\gamma)H(Z)+\gamma H(Z|X_1)+\gamma H(Z|X_2),\IEEEeqnarraynumspace\IEEEyesnumber\label{eq:fg_sto_sep}
\end{IEEEeqnarray}
where the subscript indicates the trade-off parameter $\gamma$. 
\begin{figure*}[t]
   \centering
   %\subfloat[Noise-Free Dataset, $|Z|=4,|X_1|=|X_2|=4$.]{%
   %  \label{subfig:y2b2_noncorr_compare}
    % \includegraphics[width=3in]{figures/toy_y2b2_nzall_vs.eps}}
   \subfloat[Noise-Free Dataset, $|\mathcal{Z}|=8,|\mathcal{X}_1|=|\mathcal{X}_2|=16$.]{%
     \label{subfig:y8b2_noncorr_compare}
     \includegraphics[width=2.9in]{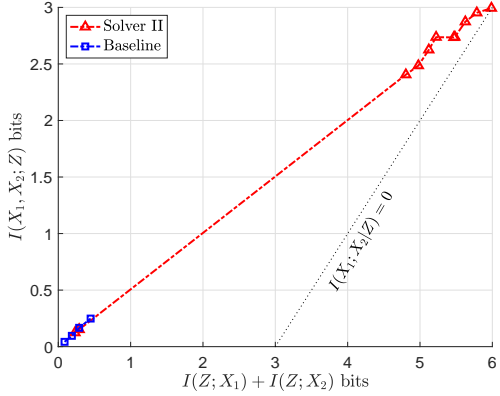}} 
   \hfil
   \subfloat[Noisy Dataset, $|\mathcal{Z}|=8,|\mathcal{X}_1|=|\mathcal{X}_2|=16$.]{%
     \label{subfig:y8b2_corr_compare}
     \includegraphics[width=2.9in]{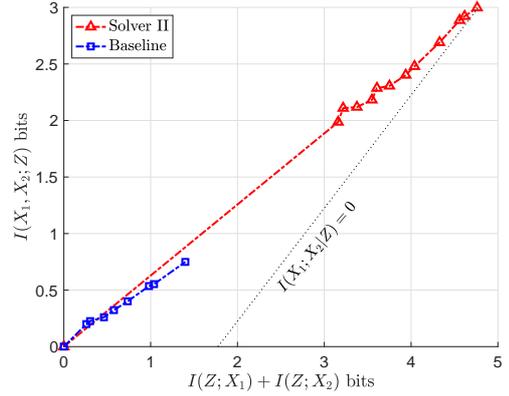}}  
   \caption{\textit{Solver \RomanNum{2}} versus the baseline method \cite{disGrad21}. Evaluated on the synthetic datasets.}
   \label{fig:y8b2_compare_baseline_all}
\end{figure*}
\begin{figure*}[t]
    \centering
    \subfloat[Noise-Free Dataset,$|\mathcal{Z}|=8,|\mathcal{X}_1|=|\mathcal{X}_2|=16$]{%
      \label{subfig:sol12_noncorr_mi}
      \includegraphics[width=2.8in]{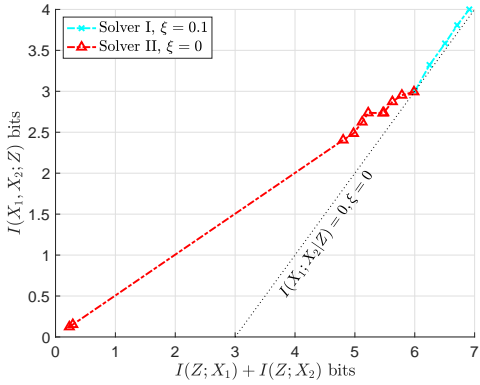}
    }
    \hfil
    \subfloat[Noisy, $|\mathcal{Z}|=8,|\mathcal{X}_1|=|\mathcal{X}_2|=16$]{%
        \label{subfig:sol12_corr_mi}
        \includegraphics[width=2.8in]{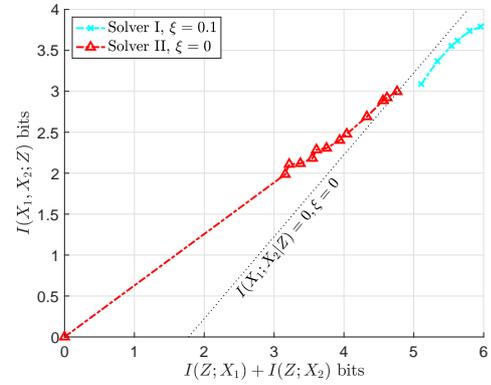}
    }
    \caption{Comparing the common information planes characterized by \textit{Solver \RomanNum{1}} and \textit{Solver \RomanNum{2}} with synthetic datasets.}
    \label{fig:sol12_compare_mi}
\end{figure*}

 Existing convergence analysis for ADMM solvers requires a convex-weakly convex structure (for the two objective functions) which is not satisfied by our formulation~\cite{themelis_patrinos_2020,huang2022admm}. To see this, one can express the negative entropy $-H(X)=-\sum_{x\in\mathcal{X}}e^{-p_x}p_x$ with $p_x:=-\log{p(x)}$. Then $-H(X)$ is convex w.r.t. the negative log-likelihood when $0\leq p_x\leq2 $ but concave when $p_x>2$. We address this issue by developing new techniques that generalize the convergence analysis to objectives with weakly convex pairs, assuming that the following conditions are satisfied:
 
 \begin{assumption}\label{assump:main_condi}
     \begin{itemize}
         \item There exists a non-empty set of stationary points $\Omega^*:=\{\omega|\nabla \mathcal{L}_c(\omega)=0,\omega \in\Omega \}$.
         \item $F$ is $\sigma_{F}$-weakly convex and $L_p$-smooth.
         \item $G$ is $\sigma_G$-weakly convex and $L_q$-smooth.
         \item The penalty coefficient $c>\max\left\{\sigma_G,{(\sigma_F+\Delta)}/{2}\right\}$, where $\Delta:=\sqrt{\sigma_F^2+8L_p^2}$
     \end{itemize}
 \end{assumption}
 The feasibility of the above conditions is based on the smoothness of the sub-objectives, which can be established by employing bounded ranges for the log-likelihoods $p,q$ (see Lemma \ref{lemma:llr_weak_cvx} and Lemma \ref{lemma:weakcvx_negent} in Appendix \ref{appexdix:conv_proof})~\cite{6203416,han2020optimal}. Our convergence result is formalized as follows
 \begin{theorem}\label{thm:main_conv}
     Let the augmented Lagrangian $\mathcal{L}_c(p,\nu,q)$ defined as in \eqref{eq:aug_lag}. Suppose Assumption \ref{assump:main_condi} holds, then the sequence $\{w^k\}_{k\in\mathbb{N}}$, where $w^k:=(p^k,\nu^k,q^k)$, obtained from the solver \eqref{eq:alg_admm_sol} converges linearly fast locally around a neighborhood of a \bf{local minima} $w^*:=(p^*,\nu^*,q^*)$.
 \end{theorem}
 \begin{IEEEproof}
     See Appendix \ref{appexdix:conv_proof}. The proof follows the outline provided by \cite[Sec. \RomanNum{3}]{huang2022admm}, but our assumptions are more relaxed since we only require weakly-convex objectives.
 \end{IEEEproof}
 The main tool for our convergence proof is the Kurdyka-{\L}ojasiewicz inequality~\cite{attouch2009convergence,li_pong_2018}, which inspired efficient solvers for Information Bottleneck~\cite{8919752,9518141,9965818,huang2022admm}. Therefore, one can view the results as extension of this earlier line of works to the Wyner multi-view unsupervised learning paradigm. 

Our results generalize to an arbitrary number of views by using the integrated Lagrangian given by:
\begin{lemma}\label{lemma:rep_lag_nv}
    The Lagrangian of the problem \eqref{eq:rep_form_gen_problem} is:
    \begin{IEEEeqnarray}{rCl}
        \mathcal{L}_{\{[\gamma]\}}'&=&\left(1+\Gamma_{[V]}\right)I(X^V;Z)\nonumber\\
        &&-\sum_{S\subset{[V]}}\gamma_S\left[I(X_S;Z)+I(X_{S^c};Z)\right],\nonumber
    \end{IEEEeqnarray}
    where $\Gamma_{[V]}:=\sum_{S\subset[V]}\gamma_S$ and $S\subset[V]$ is a partition of the indices set as in \eqref{eq:rep_form_gen_problem}.
\end{lemma}
\begin{IEEEproof}
    See Appendix \ref{appendix:multiview_rep_form}.
\end{IEEEproof}

Finally, two remarks are in order:

\begin{enumerate}
\item By setting $I(X_1;X_2|Z)=0$ in \eqref{eq:condmi_expand}, a linear equation of the two objectives is revealed:
\begin{equation}\label{eq:sto_line}
    I(Z;X_1,X_2)=\sum_{i=1}^2I(Z;X_i)-I(X_1;X_2),
\end{equation}
which allows for comparing different solvers in terms of the characterization of a fundamental trade-off as shown in Section \ref{Sec:Eval}.
\item Contrary to the variational form solver, here we cannot guarantee a linearly growing complexity with the number of views. The characterization of such scaling behavior remains an open problem.  
\end{enumerate}

\begin{figure*}[t]
   \centering
   \subfloat[Noise-Free Dataset.]{%
     \label{subfig:test_cr0_toyy8b2}
     \includegraphics[width=2.9in]{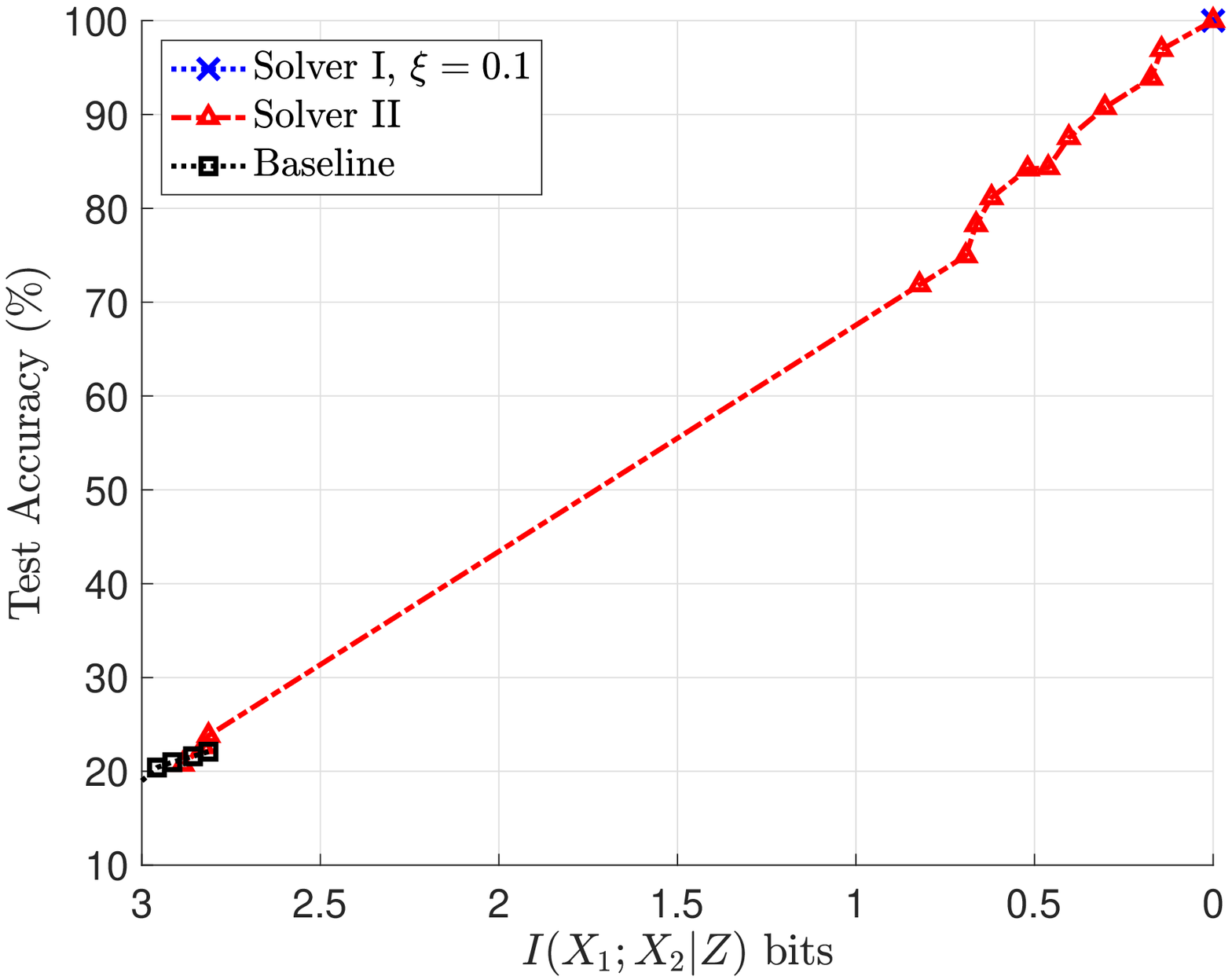}}  
   \hfil
   \subfloat[Noisy Dataset.]{%
     \label{subfig:test_cr02_toyy8b2}
     \includegraphics[width=2.8in]{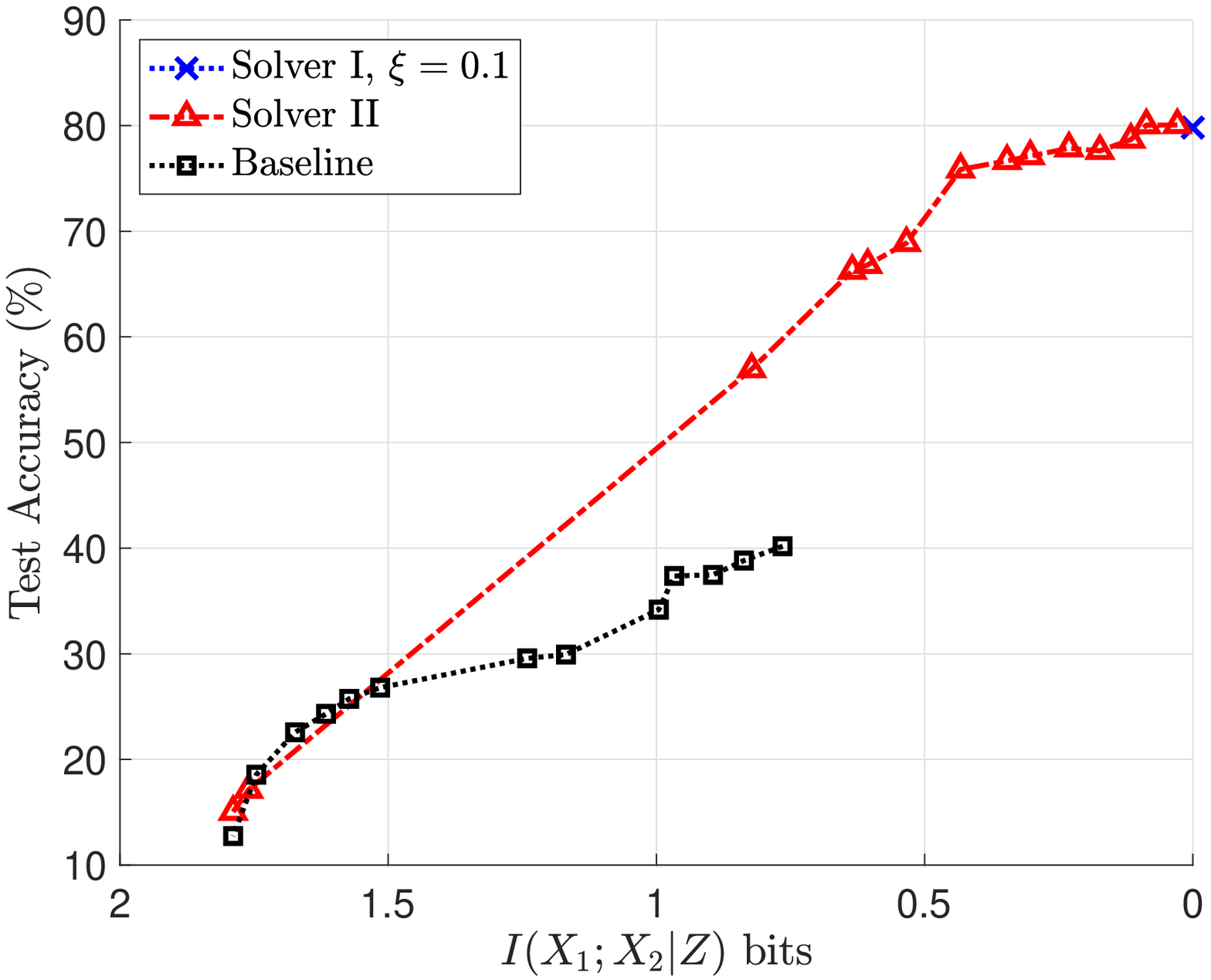}}  
   \caption{Testing accuracy versus conditional mutual information. Only one point for Solver I is reported here, corresponding to $I(X_1,X_2|Z)\approx 0$.}
   \label{fig:test_accuracy_toy}
\end{figure*}

\section{Numerical Results}\label{Sec:Eval}
In this section, we evaluate \textit{Solver \RomanNum{1}} and \textit{Solver \RomanNum{2}} and compare their performance to the gradient descent-based solver~\cite{disGrad21} (denoted as \textit{Baseline}). We use synthetic two-view datasets where the joint distribution is available but not the ground-truth, i.e., $P(Y),P(X_1|Y),P(X_2|Y)$ in the distribution $P(X_1,X_2)=\sum_{y\in\mathcal{Y}}P(y)P(X_1|y)P(X_2|y)$ are hidden. The synthetic datasets have $P(y)=1/|\mathcal{Y}|,\forall y\in\mathcal{Y}$. 
%The first dataset has:
%\begin{equation} 
%    P^{(1)}(X_i|Z)=\begin{bmatrix}
%    0.5 & 0\\
%    0.5 & 0\\
%    0 & 0.5\\
%    0 & 0.5
%    \end{bmatrix},\forall i\in{1,2},
%\end{equation}
%and $P^{(1)}(X_1,X_2)=\sum_{z\in\mathcal{Z}}p(z)p^{(1)}(z|X_1)p^{(1)}(z|X_2)$.
with the conditional probability for each view:
%\begin{equation}
%    \begin{split}
%    {}&P^{(2)}(X_i|Z)=\\
%    &\begin{bmatrix}
%    \frac{1}{2}-\delta& 0 & 0 & 0 & 0 & 0 & 0 &\delta\\
%    \frac{1}{2}-\delta& 0 & 0 & 0 & 0 & 0 & 0 &\delta\\
%    \delta&\frac{1}{2}-\delta& 0 & 0 & 0 & 0  & 0 &0\\
%    \delta&\frac{1}{2}-\delta& 0 & 0 & 0 & 0  & 0 &0\\
%    0&\delta&\frac{1}{2}-\delta& 0 & 0 & 0 & 0  & 0 \\
%    0&\delta&\frac{1}{2}-\delta& 0 & 0 & 0 & 0  & 0 \\
%    0&0&\delta&\frac{1}{2}-\delta& 0 & 0 & 0 & 0  \\
%    0&0&\delta&\frac{1}{2}-\delta& 0 & 0 & 0 & 0  \\
%    0&0&0&\delta&\frac{1}{2}-\delta& 0 & 0 & 0  \\
%    0&0&0&\delta&\frac{1}{2}-\delta& 0 & 0 & 0  \\
%    0&0&0&0&\delta&\frac{1}{2}-\delta& 0 & 0   \\
%    0&0&0&0&\delta&\frac{1}{2}-\delta& 0 & 0   \\
%    0&0&0&0&0&\delta&\frac{1}{2}-\delta& 0   \\
%    0&0&0&0&0&\delta&\frac{1}{2}-\delta& 0   \\
%    0&0&0&0&0&0&\delta&\frac{1}{2}-\delta \\
%    0&0&0&0&0&0&\delta&\frac{1}{2}-\delta
%    \end{bmatrix}
%    \end{split}
%\end{equation}

\begin{equation}\label{eq:eval_dataset}
\begin{split}
P(X_i|Y)=
    &\begin{bmatrix}
        \boldsymbol{\frac{1}{2}-\delta} &\boldsymbol{0}&\cdots&\boldsymbol{0}&\boldsymbol{\delta}\\
        \boldsymbol{\delta}& \boldsymbol{\frac{1}{2}-\delta}&\boldsymbol{0}&\dots &\boldsymbol{0}\\
        \boldsymbol{0} & \boldsymbol{\delta}&\boldsymbol{\frac{1}{2}-\delta} & \ddots&\vdots \\
        \vdots & \ddots &\ddots & \ddots&\boldsymbol{0}\\
        \boldsymbol{0} & \cdots&\boldsymbol{0}&\boldsymbol{\delta} &\boldsymbol{\frac{1}{2}-\delta}
    \end{bmatrix},
\end{split}
\end{equation}
where each bold-face symbol $\boldsymbol{v}$ is a vector $v\cdot\boldsymbol{1},v\in\mathbb{R}$ with $\boldsymbol{1}$ denotes the all-one vector. We set $\boldsymbol{1}\in\mathbb{R}^2$ (e.g., $\mathbf{\frac{1}{2}}=\begin{bmatrix}\frac{1}{2}&\frac{1}{2}\end{bmatrix}^T$), $|\mathcal{Y}|=8$ and $|\mathcal{X}_1|=|\mathcal{X}_2|=16$. We consider two cases of \eqref{eq:eval_dataset}: 1) \textit{Noise-Free} $\delta=0$ and 2) \textit{Noisy} $\delta=0.1$. Note that we assume knowledge of $|\mathcal{Z}|=|\mathcal{Y}|$. 
Both \textit{Solver \RomanNum{1}} and
\textit{Solver \RomanNum{2}} are implemented with gradient-descent with a fixed step-size $10^{-2}$. The trade-off parameter for \textit{Solver \RomanNum{1}} is set to $\beta=6.0$, and the penalty coefficient for \textit{Solver \RomanNum{2}} is $c=128$. \textit{Solver \RomanNum{1}} is initialized by normalizing uniformly distributed random samples for a pair of $(P(X_1|Z),P(X_2|Z))$, and hence $(q_{x_1|z},q_{x_2|z})$. \textit{Solver \RomanNum{2}} and \textit{Baseline} are initialized similarly.

We first compare \textit{Solver \RomanNum{2}} and \textit{Baseline} as both solvers minimize the Lagrangian \eqref{eq:mv_sto_lag}. The \textit{Baseline} is as implemented in \cite[Algorithm 1]{disGrad21}. We set the range of the trade-off parameter $\gamma\in[1,50]$ with $20$ geometrically-spaced grid points. $10$ trials are performed for each grid point. The termination criterion is $\lVert p^k-q^k\rVert^2\leq 2\times 10^{-6},\forall k\leq K_{max}=3\times 10^{5}$ for \textit{convergence}. Otherwise \textit{divergence} as a result. 
We report the converged solutions and plot them on the information plane to demonstrate the trade-off. The plane is formed by the two competing objectives: $I(Z;X_1,X_2)$ as the $y$-axis and $I(Z;X_1)+I(Z;X_2)$ as the $x$-axis. In Fig. \ref{fig:y8b2_compare_baseline_all}, the solutions from the proposed solver better characterize the trade-off and clearly outperform the \textit{Baseline} in both the \textit{Noise-Free} and \textit{Noisy} datasets, where the solutions from our solver obtain the Wyner common information (the intersection with the auxiliary dotted line \eqref{eq:sto_line}) in contrast to the \textit{Baseline} which fails to attain this point.

Next we compare \textit{Solver \RomanNum{1}} and \textit{Solver \RomanNum{2}} on the information planes. \textit{Solver \RomanNum{1}} minimizes \eqref{eq:nv_vi_lag} with $V=2,\xi\leq0.1$ and vary $|\mathcal{Z}|\in\{8,\cdots,16\}$. For \textit{Solver \RomanNum{1}}, the convergence criterion is set to $\lVert q^{k+1}_{x_i|z}-q^{k}_{x_i|z}\rVert\leq 2\times 10^{-6},\forall i\in\{1,2\}$. \textit{Solver \RomanNum{2}} follows the same configuration mentioned previously. In Fig. \ref{fig:sol12_compare_mi}, we report the minimum $I(X_1,X_2;Z)$ achieved at a fixed $I(X_1;Z)+I(X_2;Z)$. The desired solution is located at $I(X_1,X_2;Z)=3$ bits such that $I(X_1;X_2|Z)=0$. As shown in Fig. \ref{subfig:sol12_noncorr_mi} and Fig. \ref{subfig:sol12_corr_mi}, in both \textit{Noise-Free} and \textit{Noisy} datasets, the two solvers attain the desired solutions with low approximation error. Note that for both solvers, the cardinality $|\mathcal{Z}^*|$ that achieves the desired solution coincides with that of the labels, i.e., $|\mathcal{Z}^*|=|\mathcal{Y}|=8$.

Finally, we apply the proposed solvers for unsupervised clustering with the \textit{Noise-Free} and \textit{Noise} datasets. Different from standard settings, we consider multi-view scenarios where all view-observations share a common target (the clusters), this additional knowledge is the key for efficient unsupervised learning with Wyner common information. We generate training and testing data as follows. $N$ samples of the labels $Y$ are generated with $P(Y)=1/|\mathcal{Y}|$. Given the labels, the observations of each view are generated from inverse transform sampling~\cite{DevroyeLuc2013NRVG} with $P(X_i|Y),\forall i\in\{1,2\}$ defined in \eqref{eq:eval_dataset}. We follow this procedure to generate $N_{\text{train}}=10^{5}$ training and $N_{\text{test}}=10^4$ data. At training phase, all methods use the estimated empirical joint distribution $P(X_1,X_2)$, obtained through counting the pairs of $(x_1,x_2)\in D_{train}$ and obtain a common information encoder $P(Z|X_1,X_2)$. Then the encoder is tested with testing data. For each test sample $(x_1,x_2)\in D_{test}$, a hypothesis $\hat{z}\in\mathcal{Z}$ is obtained through inverse transform sampling with $P(Z|x_1,x_2)$. After collecting all $\{\hat{z}\}$, exhaustive search is adopted to find the best labeling over all permutations of $\mathcal{Z}\mapsto\mathcal{Y}$ (accomplished through one shot training)~\cite{xie2016unsupervised,Federici2020Learning,tian2020contrastive}, and we obtain the testing accuracy accordingly. In Fig. \ref{fig:test_accuracy_toy} we report the testing accuracy achieved by the two new solvers and \textit{Baseline} for both \textit{Noise-Free} and \textit{Noisy} datasets. Clearly, the proposed solvers outperform the \textit{Baseline} over a wide range of $I(X_1;X_2|Z)$. Notably, our solvers attain the Wyner common information $I(X_1;X_2|Z)\approx 0$ with $>98\%$ accuracy in \textit{Noise-Free} and $>78\%$ accuracy in \textit{Noisy} datasets in contrast to the \textit{Baseline}.

\section{Conclusion}
\label{sec:conclusion}

In this work, we proposed two forms for Discrete Wyner multi-view unsupervised learning leading to two computationally efficient solvers with guaranteed convergence to local minima. The first borrows tools from variational inference and is shown to enjoy a linear complexity scaling with the number of views. The second solver is obtained by adapting the ADMM to a representational learning form of Wyner common information. The convergence analysis of the ADMM-based solver is shown to generalize earlier works on the Information Bottleneck solvers to our context~\cite{9518141,huang2022admm}. Numerically, both solvers are shown to outperform the existing gradient descent-based approach on a synthetic dataset. While our focus here was devoted to the discrete case, we have generalized our variational approach to continuous distributions and the corresponding excellent results are reported elsewhere. Overall, both theoretical and numerical results demonstrate the efficacy of the proposed Wyner solvers for multi-view unsupervised learning.

%%%%%%
%% To balance the columns at the last page of the paper use this
%% command somewhere at the top of the first column of the last page:
%%
% \enlargethispage{-5cm} 
%%
%% where the exact amount of page reduction has to be adapted to the
%% actual situation.
%%
%% If the balancing should occur in the middle of the references, use
%% the following trigger:
%%
% \IEEEtriggeratref{3}
%%
%% which triggers a \newpage (i.e., new column) just before the given
%% reference number. Note that you need to adapt this if you modify
%% the paper. The "triggered" command can be changed if desired:
%%
% \IEEEtriggercmd{\enlargethispage{-20cm}}
%%
%%%%%%

%%%%%%
%% References:
%% We recommend the usage of BibTeX:
%%
\bibliographystyle{IEEEtran}
\bibliography{references}
%%
%% where we here have assume the existence of the files
%% definitions.bib and bibliofile.bib.
%% BibTeX documentation can be obtained at:
%% http://www.ctan.org/tex-archive/biblio/bibtex/contrib/doc/
%%%%%%
%% Or you use manual references (pay attention to consistency and the
%% formatting style!):

%\begin{thebibliography}{9}

%\bibitem{Laport:LaTeX}
%L.~Lamport,
%  \emph{\LaTeX: A Document Preparation System,} 
%  Addison-Wesley, Reading, Massachusetts, USA, 2nd~ed., 1994. 

%\bibitem{GMS:LaTeXComp}
%F.~Mittelbach, M,~Goossens, J.~Braams, D.~Carlisle, and
%C.~Rowley, \emph{The {\LaTeX} Companion,} Addison-Wesley,
%Reading, Massachusetts, USA, 2nd~ed., 2004.

%\bibitem{oetiker_latex}
%T.~Oetiker, H.~Partl, I.~Hyna, and E.~Schlegl, \emph{The Not So Short
%  Introduction to {\LaTeX2e}}, version 6.4, Mar.~9, 2021. [Online].
%  Available: \url{https://tobi.oetiker.ch/lshort/}

%\bibitem{typesetmoser}
%S.~M. Moser, \emph{How to Typeset Equations in {\LaTeX}}, version 4.6,
%  Sep. 29, 2017. [Online]. Available:
%  \url{https://moser-isi.ethz.ch/manuals.html#eqlatex}

%\bibitem{shell15}
%M.~Shell, ``How to use the {IEEEtran} {\LaTeX} class,'' \emph{Journal of
%  {\LaTeX} Class Files}, vol.~14, no.~8, Aug. 2015. [Online]. Available:
%  \url{https://mirrors.ctan.org/macros/latex/contrib/IEEEtran/IEEEtran\_HOWTO.pdf}

%\bibitem{IEEE:AuthorToolbox}
%IEEE, \emph{Author Center.} [Online.] Available:
%  \url{https://ieeeauthorcenter.ieee.org/}

%\end{thebibliography}

%%%%%% 
%% Appendix:
%% If needed a single appendix is created by
%%
%\appendix
%%
%% If several appendices are needed, then the command
%%
 \appendices
%%
%% in combination with further \section-commands can be used.
%%%%%%
\section{Proof of Lemma \ref{lemma:gen_vi_bound}}\label{appendix:pf_gen_vi_bound}
The proof follows by the following derivations:
\begin{IEEEeqnarray}{rCl}\label{eq:gennv_vi_bound}
    &{}&I_\theta(X^V;Z)\nonumber\\
    &=&\mathbb{E}_{Z}\left[D_{KL}[P_\theta(X^V|Z)\parallel \sum_{z\in\mathcal{Z}}P_\theta(z)P_\theta(X^V|z)]\right]\nonumber\\
    &=&\mathbb{E}_{Z}\left[\sum_{x^V\in\mathcal{X}^V}P_\theta(x^V|Z)\log{\frac{P_\theta(x^V|Z)}{\sum_{z'\in\mathcal{Z}}P_\theta(z')P_\theta(x^V|z')}}\right]\nonumber\\
    &=&-H_\theta(X^V|Z)-\mathbb{E}_{X^V;\theta}\left[\log{\sum_{z'\in\mathcal{Z}}P_\theta(z')P_\theta(X^V|z')}\right]\nonumber\\
    &=&-H_\theta(X^V|Z)\nonumber\\
    &&-\mathbb{E}_{X^V;\theta}\left[\log{\sum_{z'\in\mathcal{Z}}P_\theta(z')P_\theta(X^V|z')}\frac{P(X^V)}{P(X^V)}\right]\nonumber\\
    &=&-\sum_{i=1}^VH_\theta(X_i|Z)-D_{KL}\left[P_\theta(X^V)\parallel P(X^V)\right]\nonumber\\
    &&-\sum_{x^V\in\mathcal{X}^V}\left[\sum_{z\in\mathcal{Z}}P_\theta(z)\prod_{i=1}^VP_\theta(x_i|z)\right]\log{P(x^V)}\nonumber\\
    &\leq&-\sum_{i=1}^VH_\theta(X_i|Z)-\mathbb{E}_{X^V;\theta}\left[\log{P(X^V)}\right],\IEEEeqnarraynumspace\IEEEyesnumber
\end{IEEEeqnarray}
where the last equality is due to the Markov chain $X_S\rightarrow Z\rightarrow X_{S^c}$ for all partitions $S$ of $[V]$, and the last inequality follows by the non-negativity of KL divergence. 

\section{Convergence Analysis}\label{appexdix:conv_proof}
In this part, we provide the convergence analysis of the proposed solver (Theorem \ref{thm:main_conv}). The analysis is based on the convergence in non-convex splitting methods. Here we focus on objective functions satisfying Assumption \ref{assump:main_condi}, but we note that the convergence analysis applies to a broader class of objectives, we refer to \cite{attouch2009convergence,attouch2010proximal,bolte2018first,bolte2018nonconvex} and the references therein for more general functions. 
\subsection{Preliminaries}\label{appendix:conv_prelim}
\begin{definition}
A function $f:\mathbb{R}^n\mapsto [0,\infty)$, with distinct $x,y\in \Omega$ is \textit{Lipschitz continuous} if:
 \begin{equation*}
 |f(x)-f(y)|\leq M|x-y|,
 \end{equation*}
where $M>0$ is the Lipschitz coefficient, and $\Omega$ denotes the feasible set through out this section if not further specify.
\end{definition}
\begin{definition}
A function $f:\mathbb{R}^n\mapsto [0,\infty),f\in\mathcal{C}^1$ is $L$-\textit{Lipschitz smooth} if:
 \begin{equation*}
 |\nabla f(x)-\nabla f(y)|\leq L|x-y|,
 \end{equation*}
$\forall x,y\in\Omega$ and $L>0$ is the smoothness coefficient.
\end{definition}
\begin{definition}
A function $f:\mathbb{R}^n\mapsto [0,\infty),f\in\mathcal{C}^1$ is $\sigma$-\textit{hypoconvex}, $\sigma\in\mathbb{R}$ if the following holds:
\begin{equation}\label{eq:hypoconvex}
    f(y)\geq f(x)+\langle\nabla f(x),y-x\rangle+\frac{\sigma}{2}\lVert y-x\rVert^2.
\end{equation}
\end{definition}
In the above definition, $\sigma=0$ reduces to the standard convexity; $\sigma>0$ is known as strong convexity whereas $\sigma<0$ is weak convexity.

In~\cite[Definition 1]{huang2022admm}, it is shown that for range-bounded probability mass functions, the (conditional) entropy function is weakly convex. Here we extend this smoothness condition to the case with log-likelihood as the variables.
\begin{lemma}\label{lemma:llr_weak_cvx}
    Let $q:=\log{p_x}\in\mathbb{R}^{|\mathcal{X}|}_{-}, \boldsymbol{1}^Tp_x=1$. Suppose $q_x<-\epsilon_0$, with $2>\epsilon_0>0$, then the entropy $H(X)$ is $\sigma$-weakly convex, where $\sigma=2e^{-\epsilon_0}(2-\epsilon_0)$.
\end{lemma}
\begin{IEEEproof}
    Since the entropy $H(X)$ can be expressed as $H(X)=-{e^{q}}^Tq,q\in\mathbb{R}^{|\mathcal{X}|}_{-}$. The Hessian of $H(X)$ w.r.t. $q$ is:
    \begin{IEEEeqnarray*}{rCl}
        \nabla_q^2H(X)=-\text{diag}\left(\text{diag}(e^{q})q\right)-2\textit{diag}(e^q).
    \end{IEEEeqnarray*}
    Therefore $z^T\nabla^2H(X)z=-\sum_{i=1}^{|\mathcal{X}|}z_i^2e^{q_i}(2+q_i)$, and without loss of generality, we can focus on $-2<q_1<-\epsilon_0$ (convex otherwise). It is straightforward to see that the negative scalar for each squared term $z_i^2$ is bounded from below by $-e^{-\epsilon_0}(2-\epsilon_0)$. We conclude that $H(X)$ is $2e^{-\epsilon_0}(2-\epsilon_0)$-weakly convex. 
\end{IEEEproof}
It is well-known that the negative entropy $H(X)$ is convex w.r.t. the $p_x$. The log-likelihood minimization technique also changes the convexity property when compared with assigning the probability mass as the variables.
\begin{lemma}\label{lemma:weakcvx_negent}
    Let $q$ be defined as in Lemma \ref{lemma:llr_weak_cvx}, then the negative entropy $-H(X)$ is $\sigma$-weakly convex, where $\sigma=2e^{-3}$.
\end{lemma}
\begin{IEEEproof}
    Using $q$, the negative entropy can be written as:$-H(X)={e^{q}}^Tq$. Following this, the Hessian matrix is:
    \begin{IEEEeqnarray*}{rCl}
        -\nabla^2_qH(X)=\text{diag}\left(\text{diag}(e^{q})q\right)+2\text{diag}(e^{q}).
    \end{IEEEeqnarray*}
Then without loss of generality, we can focus on the first element within $q_1<-2$ (convex otherwise). In this range, the entry $[-\nabla^2_qH(X)]_{(1,1)}$ is lower bounded by $-e^{-3}$. Therefore, we conclude that $-H(X)$ is $2e^{-3}$-weakly convex.
\end{IEEEproof}

The log-likelihood optimization technique facilitates the establishment of smoothness condition. It is known that the smoothness condition depends on the minimum non-zero element of a probability mass~\cite{han2020optimal,6203416,huang2022admm}. Then consider a range bounded measure $\mu,\sum_{x\in\mathcal{X}}\mu(x)=1$ such that $\epsilon<\mu(x)<1-\epsilon,\forall x\in\mathcal{X},0<\epsilon< 1$. In contrast, $-\infty<-M_{\epsilon}<\log{\mu(x)}<-m_{\epsilon}$, where $M_\epsilon$ is the largest machine number, $m_\epsilon$, the machine precision. Clearly, $M_\epsilon$ can be significantly larger than $1-\epsilon$ in terms of machine precision. In practice, the log-likelihood optimization technique enables more relaxed simplex projection as the range of simplex expands from $(\epsilon,1-\epsilon)$ to $(-M_\epsilon,-m_\epsilon)$. 

\subsection{Convergence Proof for Solver \RomanNum{1}}\label{appendix:conv_solver1}
The overall Lagrangian \eqref{eq:nv_vi_lag} is minimized alternatingly with \eqref{eq:vi_alg}. Therefore, it suffices to consider a single update. Without loss of generality, we focus on the update for the first view $q_{x_1|z}$, i.e., the vector form of $P_\theta(X_1|Z)$:
\begin{IEEEeqnarray}{rCl}
    \mathcal{L}^{(1)}_{\beta}:&=&-H_\theta(X_1|Z)+\beta\sum_{x^V\in\mathcal{X}^V}P(x^V)\left[\log{P(x^V)}\right.\nonumber\\
    &&\left.\vphantom{\log{P(x^V)}}-\log{\sum_{z'\in\mathcal{Z}}\frac{1}{|\mathcal{Z}|}P_\theta(X_1|z')\prod_{j\neq 1}P_\theta(X^k_j|z')}\right].\IEEEeqnarraynumspace\IEEEyesnumber\label{eq:appendix_conv1_single1}
\end{IEEEeqnarray}

The first term in the above is a negative (conditional) entropy, which is convex w.r.t. $P_\theta(X_1|Z)$ since $P_\theta(Z)=1/|\mathcal{Z}|$ \cite{cover1999elements}. The second term is a constant that is independent to $P_\theta(X_1|Z)$ and the last term is convex w.r.t. $P_\theta(X^V)$, and $P_\theta(X^V)$ is an affine transform of $P_\theta(X_1|Z)$ (since $\prod_{j\neq 1}P_\theta(X_j^k|Z)$ is fixed). Therefore the last term in \eqref{eq:appendix_conv1_single1} is convex w.r.t. $P_\theta(X_1|Z)$. Since \eqref{eq:appendix_conv1_single1} is convex w.r.t. $P_\theta(X_1|Z)$, and similarly for all $i\in[V]$, the convergence is guaranteed from elementary convex optimization analysis~\cite{nesterov2003introductory,NesterovYurii2018Loco,BertsekasDimitriP.1999Np,boyd2004convex}. In other words, each step update of \eqref{eq:vi_alg} minimizes a convex sub-objective over $P_\theta(X_i|Z),\forall i\in[V]$ (by fixing $\{P_\theta(X_j|Z)\}_{j\neq i}$) which assures convergence but since the full objective function is non-convex w.r.t. $\{P_\theta(X_j|Z)\}_{j=1}^V$ hence the converged solution is qualified to a local minima.

\subsection{Convergence Proof for Solver \RomanNum{2}}\label{appendix:conv_solver2}
In this part, we prove the convergence of the proposed solver \eqref{eq:alg_admm_sol} for the non-linearly constrained augmented Lagrangian \eqref{eq:aug_lag}. Throughout this part, we assume that Assumption \ref{assump:main_condi} holds. For simplicity of expression, we denote $\mathcal{L}_c^k:=\mathcal{L}_c(p^k,\nu^k,q^k)$ as the augmented Lagrangian \eqref{eq:aug_lag} evaluated with a step $k$ collective point $w^k:=(p^k,\nu^k,q^k)$, and the subscript for the sub-objectives $F_s$ are omitted for simplicity of expression.

For the proposed solver \eqref{eq:alg_admm_sol}, the first-order necessary conditions are as follows:
\begin{IEEEeqnarray}{rCl}\label{eq:solver2_min_fonc}
0&=&\nabla F(p^{k+1})+\nu^k+c\left(p^{k+1}-q^k\right)\nonumber\\
&=&\nabla F(p^{k+1})+\nu^{k+1},\IEEEyessubnumber\\
\nu^{k+1}&=&\nu^k+c\left(p^{k+1}-q^k\right),\IEEEyessubnumber\\
0&=&\nabla G(q^{k+1})-\left[\nu^{k+1}+c\left(p^{k+1}-q^{k+1}\right)\right].\IEEEyessubnumber
\end{IEEEeqnarray}
And the main theorem is re-stated here for completeness.
\begin{customthm}{1}
Let $\mathcal{L}_c$ be defined as in \eqref{eq:aug_lag}. Suppose Assumption \ref{assump:main_condi} is satisfied. If the sequence $\{w^k\}_{k\in\mathbb{N}}$ is obtained through \eqref{eq:alg_admm_sol}, where $w_k:=(p^k,\nu^k,q^k)$ denotes the collective point at step $k$, then $\{w^k\}$ converges to a stationary point $w^*\in\Omega^*$ at linear rate.
\end{customthm}
\begin{IEEEproof}
The proof is based on the sufficient decrease lemma (Lemma \ref{lemma:suf_dec_solver2}). Then the convergence and locally rate of convergence results follow by \cite[Theorem 2]{huang2022admm} with the assignment of variables as $p:=-\log{ p_{z|x_1,x_2}}=q$, and the linear operators $A=B=I$.
\end{IEEEproof}

%% RULES%%%%%%%
%The appendix (or appendices) are optional. For reviewing purposes
%additional 5~pages (double-column) are allowed (resulting in a maximum
%grand total of 10~pages plus one page containing only
%references). These additional 5~pages must be removed in the final
%version of an accepted paper.
\begin{lemma}\label{lemma:suf_dec_solver2}
Under Assumption \ref{assump:main_condi}, if the sequence $\{w^k\}_{k\in\mathbb{N}}$ is obtained through \eqref{eq:alg_admm_sol}, then we have:
\begin{equation*}
    \mathcal{L}^k-\mathcal{L}^{k+1}\geq \delta_p\lVert p^k-p^{k+1}\rVert^2+\delta_q \lVert q^k-q^{k+1}\rVert^2,
\end{equation*}
for some non-negative coefficients:
\begin{equation*}
    \delta_p:= \frac{c-\sigma_F}{2}-\frac{L_p^2}{c},\,\delta_q:=\frac{c-\sigma_G}{2}.
\end{equation*}
\end{lemma}
\begin{IEEEproof}
    Expand the sub-objective according to the solver \eqref{eq:alg_admm_sol}, the proof can be divided into three parts:
    \begin{subequations}
        \begin{align}
        {}&\mathcal{L}_c(p^k,\nu^k,q^k)-\mathcal{L}_c(p^{k+1},\nu^{k+1},q^{k+1})\nonumber\\
        =&\mathcal{L}_c(p^k,\nu^k,q^k)-\mathcal{L}_c(p^{k+1},\nu^k,q^k)\label{subeq:lem_p_step}\\
        &+\mathcal{L}_c(p^{k+1},\nu^k,q^k)-\mathcal{L}_c(p^{k+1},\nu^{k+1},q^k)\label{subeq:lem_nu_step}\\
        &+\mathcal{L}_c(p^{k+1},\nu^{k+1},q^k)-\mathcal{L}_c(p^{k+1},\nu^{k+1},q^{k+1}).\label{subeq:lem_q_step}
        \end{align}
    \end{subequations}
For \eqref{subeq:lem_p_step}, we have the follow lower bound:
\begin{IEEEeqnarray*}{rCl}
    {}&{}&\mathcal{L}_c(p^k,\nu^k,q^k) - \mathcal{L}_c(p^{k+1},\nu^k,q^k)\\
    &=&F(p^{k})-F(p^{k+1})+\langle \nu^{k},p^k-p^{k+1}\rangle+\frac{c}{2}\lVert p^{k}-q^k\rVert^2\\
    &&-\>\frac{c}{2}\lVert p^{k+1}-q^k\rVert^2\\
    &\geq&\langle\nabla F(p^{k+1})+\nu^k,p^k-p^{k+1} \rangle-\frac{\sigma_{F}}{2}\lVert p^k-p^{k+1}\rVert^2\\
    && +\>\frac{c}{2}\lVert p^k-q^k\rVert^2-\frac{c}{2}\lVert p^{k+1}-q^k\rVert^2\\
    &=&-c\langle p^{k+1}-q^{k},p^k-p^{k+1} \rangle-\frac{\sigma_F}{2}\lVert p^k-p^{k+1}\rVert^2\\
    &&+\>\frac{c}{2}\lVert p^k-q^k\rVert^2-\frac{c}{2}\lVert p^{k+1}-q^k\rVert^2\\
    &=&\frac{c-\sigma_{F}}{2}\lVert p^k-p^{k+1}\rVert^2,
\end{IEEEeqnarray*}
where the first inequality is due to the $\sigma_F$-weak convexity of $F$ and the last equality is due to \eqref{eq:solver2_min_fonc} along with the relation:
\begin{equation}\label{eq:three_point}
    2\langle u-w,v-u \rangle= |w-v|-|u-w|-|u-v|.
\end{equation}
%From \eqref{eq:sol2_suf_f_ineq}, the $\sigma_F$-weak convexity of $F$ can be balanced by letting $c>\sigma_F$.
As for \eqref{subeq:lem_nu_step}, we have:
\begin{equation}\label{eq:pf_lem_suf_nu}
    \mathcal{L}_c(\nu^k)-\mathcal{L}_c(\nu^{k+1})=-\frac{1}{c}\lVert \nu^k-\nu^{k+1}\rVert^2,
\end{equation}
where we denote $\mathcal{L}_c(\nu):=\mathcal{L}_c(p^{k+1},\nu,q^k)$. 

To find a lower of \eqref{eq:pf_lem_suf_nu}, consider the $L_p$-smoothness of the sub-objective $F$ and \eqref{eq:solver2_min_fonc}:
\begin{IEEEeqnarray*}{rCl}
    \lVert \nabla F^{k+1}-\nabla F^k\rVert&=&\lVert \nu^{k+1}-\nu^k\rVert
    \leq L_p\lVert p^{k+1}-p^{k}\rVert,\IEEEeqnarraynumspace
\end{IEEEeqnarray*}
where we denote $\nabla F^k:=\nabla F(p^k)$. Substute the above into \eqref{eq:pf_lem_suf_nu}, we get:
\begin{equation}
    -\frac{1}{c}\lVert \nu^k-\nu^{k+1}\rVert^2\geq - \frac{L_p^2}{c}\lVert p^k-p^{k+1}\rVert^2.
\end{equation}
Finally, for \eqref{subeq:lem_q_step}, we have:
\begin{IEEEeqnarray*}{rCl}
    &{}&\mathcal{L}_c(p^{k+1},\nu^{k+1},q^k)-\mathcal{L}_c(p^{k+1},\nu^{k+1},q^{k+1})\\
    &=&G(q^k)-G(q^{k+1})+\langle\nu^{k+1},q^{k+1}-q^k \rangle+\frac{c}{2}\lVert p^{k+1}-q^k\rVert^2\\
    &&-\>\frac{c}{2}\lVert p^{k+1}-q^{k+1}\rVert^2\\
    &\geq&\langle\nabla G(q^{k+1})-\nu^{k+1},q^k-q^{k+1} \rangle-\frac{\sigma_{G}}{2}\lVert q^k-q^{k+1}\rVert^2\\
    &&+\>\frac{c}{2}\lVert p^{k+1}-q^k\rVert^2-\frac{c}{2}\lVert p^{k+1}-q^{k+1}\rVert^2\\
    &=&c\langle p^{k+1}-q^{k+1},q^k-q^{k+1} \rangle-\frac{\sigma_G}{2}\lVert q^k-q^{k+1}\rVert^2\\
    &&+\>\frac{c}{2}\lVert p^{k+1}-q^k\rVert^2-\frac{c}{2}\lVert p^{k+1}-q^{k+1}\rVert^2\\
    &=&\frac{c-\sigma_G}{2}\lVert q^k-q^{k+1}\rVert^2,
\end{IEEEeqnarray*}
where the first inequality is due to the $\sigma_G$-weak convexity of $G$ and the last equality is due to \eqref{eq:solver2_min_fonc} along with \eqref{eq:three_point}.
Combining the above completes the proof.
\end{IEEEproof}

\section{Proof of Lemma \ref{lemma:rep_lag_nv}}\label{appendix:multiview_rep_form}
For an arbitrary number of views, say $V$ views, our goal is to show that the \textit{Representation} form can be expressed as a combination of Shannon entropy and conditional entropy functions as in the two-view special case \eqref{eq:mv_sto_lag}. 

Using Lagrange multiplier, the unconstrained form of the optimization problem \eqref{eq:rep_form_gen_problem} can be written as:
\begin{IEEEeqnarray}{rCl}
    \mathcal{L}_{\{\gamma\}}:=I(X^V;Z)+\sum_{S\subset[V]}\gamma_SI(X_S;X_{S^c}|Z).\IEEEeqnarraynumspace\IEEEyesnumber\label{eq:appendix_rep_nv_lag}
\end{IEEEeqnarray}
For the conditional mutual information, we have the following expansion for each partition $S\subset[V]$:
\begin{IEEEeqnarray}{rCl}
    I(X_S;X_{S^c}|Z)&=&I(X_S,X_{S^c};Z)-I(X_S;Z)-I(X_{S^c};Z)\nonumber\\
    &&+I(X_S;X_{S^c}).\IEEEeqnarraynumspace\IEEEyesnumber\label{eq:appendix_rep_nv_condmi}
\end{IEEEeqnarray}
The last term in \eqref{eq:appendix_rep_nv_condmi} is a constant since $P(X^V)$ is given (hence knowing all the conditional distribution $P(X_S|X_{S^c})$ and Marginals $P(X_S)$).

Substitute \eqref{eq:appendix_rep_nv_condmi} into \eqref{eq:appendix_rep_nv_lag}, omitting the constants, we have:
\begin{IEEEeqnarray}{rCl}
    \mathcal{L}'_{\{\gamma\}}&=&\left(1+\sum_{S\subset{[V]}}\gamma_S\right)I(X^V;Z)\nonumber\\
    &&-\sum_{S\subset{[V]}}\gamma_S\left[I(X_S;Z)+I(X_{S^c};Z)\right].\IEEEeqnarraynumspace\IEEEyesnumber\label{eq:rep_nv_expand_mi}
\end{IEEEeqnarray}
Then we can rewrite \eqref{eq:rep_nv_expand_mi} as a combination of entropy and conditional entropy function, which gives:
\begin{IEEEeqnarray}{rCl}
    \mathcal{L}'_{\{\gamma\}}&=&\left(1-\Gamma_{[V]}\right)H(Z)-\left(1+\Gamma_{[V]}\right)H(Z|X^V)\nonumber\\
    &&\sum_{S\subset[V]}\gamma_S\left[H(Z|X_S)+H(Z|S^c)\right],\IEEEeqnarraynumspace\IEEEyesnumber\label{eq:appendix_rep_nv_ent}
\end{IEEEeqnarray}
where $\Gamma_{[V]}:=\sum_{S\in[V]}\gamma_S>0$. Each term in \eqref{eq:appendix_rep_nv_ent} can be computed through marginalization through the common information encoder $P_\theta(Z|X^V)$ and the associated conditional distribution $P(X_{S^c}|X_S)$.
\end{document}